\documentclass[3p,preprint,11pt]{elsarticle}

\usepackage{amssymb,amsmath,amsthm,bm}
\usepackage[normalem]{ulem}
\usepackage{multirow}
\usepackage[retainorgcmds]{IEEEtrantools}
\usepackage{lineno}

\newcommand\NOT{{\rm not}}

\newtheorem{property}{Property}

\DeclareMathOperator{\dec}{dec}
\DeclareMathOperator{\eXOR}{eXOR}
\DeclareMathOperator{\XOR}{XOR}

\biboptions{sort&compress}

\usepackage{graphicx}

\newlength\figwidth
\newlength\imagewidth
\usepackage{color}
\definecolor{dgreen}{rgb}{0,.6,0}

\begin{document}
\begin{frontmatter}

\title{A chaotic image encryption scheme owning temp-value feedback}

\author[hk-cityu-ee]{Leo Yu Zhang\corref{corr}}
\ead{leoxtu@gmail.com}
\author[cn-sg-epri]{Xiaobo Hu}
\author[cn-xtu-cie]{Yuansheng Liu}
\author[hk-cityu-ee]{Kwok-Wo Wong}
\cortext[corr]{Corresponding author.}
\address[hk-cityu-ee]{Department of Electronic Engineering, City University of Hong Kong, Hong Kong, China}
\address[cn-sg-epri]{State Grid Electric Power Research Institute, Qinghe, Beijing 100192, China}
\address[cn-xtu-cie]{College of Information Engineering, Xiangtan University, Xiangtan 411105, Hunan, China}

\begin{abstract}
Many round-based chaotic image encryption algorithms employ the permutation-diffusion structure.
Schemes using this structure have been found insecure when the iteration round is equal to one
and the secret permutation of some existing ones can be recovered even a higher iteration round is adopted.
In this paper, we present a single round permutation-diffusion chaotic cipher for gray image,
in which some temp-value feedback mechanisms are introduced to resist the known attacks.
Specifically, we firstly embed the plaintext feedback technique in the permutation process to
develop different permutation sequences for different plain-images and then employ plaintext/ciphertext
feedback for diffusion to generate equivalent secret key dynamically. Experimental results show that the
suggested scheme owns large key space and can resist the differential attack. It is also efficient.

\end{abstract}

\end{frontmatter}

\section{Introduction}
In the information era, digital images have been widely used for various applications, such
as entertainment, business,
health service and military affairs, etc. All the sensitive data should be encrypted before transmission to
avoid eavesdropping. However, bulk data size and high redundancy among the raw pixels of a digital image
make the traditional encryption algorithms, such as DES, IDEA, AES, not able to be operated efficiently.
Therefore, designing specialized encryption algorithms for digital images has attracted much research effort.
Some intrinsic properties of chaotic systems, such as ergodicity, sensitive to the initial condition and control
parameters, are analogous to the confusion and diffusion properties specified by Shannon~\cite{Shannon:Entropy:BSTJ49}.
Thus makes it natural to employ chaotic systems in image encryption algorithms
\cite{Fridrich:ChaoticImageEncryption:IJBC98,Jakimoski:Blockcipher:IEEECirSys01,Chen:3DChaoticCipher:CSF04,Zhang:ImageCrypt:SCSF07,
Zhu:HyperEnc:OC12,
Francois:ImageEncChaos:SGIC12,Mao:3Dbakermap:Optic2009,Huang:pixelShuffle:Optic2009,Pareek:subdiffusion:CNSNS10}.
Meanwhile, the art of cryptanalysis has also made new achievements in the last few decades.
Some of the existing image encryption algorithms
are found insecure
 \cite{Alvarez:AnalyTable:PLA04,Solak:AnalyFridrich:IJBC10,Chengqing:OptimalofPermutation:SG11,LeoYu:AnalyofAlternate:IJBC12,
LeoYu:BreakColor:ND12,Liaoxf:AttackonTent:IEEECirSys06,Shujun:quantitativePermutation:SGIC08,chengqing:breakingPareek:CNSNS11}
to different degrees due to the following defects:
1) the (equivalent) secret key can be obtained by the brute-force attack due to the dynamical degradation
of chaotic systems in digital domain;
2) all the operations employed in the encryption process are reversible without the key or even linear, therefore the
mathematical model of the scheme is not a keyed one-way function \cite{Xuejia:definingCypto:2012}.
In \cite{shujun:BasicRrequirement:IJBC06}, some basic requirements for evaluating the chaotic encryption algorithms
are concluded.

The permutation-diffusion structure becomes the basis of many chaotic image encryption schemes
since Fridrich developed a chaos-based image encryption scheme of this structure in $1998$
\cite{Fridrich:ChaoticImageEncryption:IJBC98}.
The symmetric image encryption scheme in \cite{Chen:3DChaoticCipher:CSF04} extended the Cat map to
three-dimensional to make it suitable for permutation in space, followed by a similar diffusion construction of Fridrich's.
In $2004$, Mao et al. proposed an image encryption algorithm, where the discrete Baker map was
employed for permutation \cite{Mao:3Dbakermap:Optic2009}.  It's worth mentioning that most image encryption
schemes of this structure have to execute the permutation and diffusion procedures alternatively for several
rounds to fulfill the security requirement, which will certainly lead to some reduction in efficiency.
Nonetheless, Solak et al. proposed a chosen ciphertext attack in $2010$ by utilizing the relationship between
the pixels in the neighboring encryption rounds \cite{Solak:AnalyFridrich:IJBC10}.
This attack is efficient for Fridrich's scheme \cite{Fridrich:ChaoticImageEncryption:IJBC98}
and it can also be applied to Chen's scheme \cite{Chen:3DChaoticCipher:CSF04}.
In addition, it is reported in \cite{Chengqing:AnalysisFridrichstructure:IEEECirSys08} that the equivalent
key of several permutation-diffusion image ciphers, such as
those suggested in \cite{Fridrich:ChaoticImageEncryption:IJBC98,Chen:3DChaoticCipher:CSF04,
Mao:3Dbakermap:Optic2009,he:colorfridrichstruture:LNCS06},
can be recovered when only one encryption round is applied.

Meanwhile, image encryption algorithms having other structures have also been developed.
In $2010$, Patidar et al. suggested a substitution-diffusion structure for color image
\cite{Pareek:subdiffusion:CNSNS10}, which was attacked in \cite{chengqing:breakingPareek:CNSNS11}.
In \cite{Huang:pixelShuffle:Optic2009}, Huang et al. presented a multi-chaotic system based
permutation scheme, in which pixel positions and bits in
the individual pixel are shuffled together to achieve permutation and substitution simultaneously.
Intuitively, permutation-only schemes are not secure against known/chosen plaintext attack. In \cite{Chengqing:OptimalofPermutation:SG11,Shujun:quantitativePermutation:SGIC08},
Li et al. proposed the quantitative and optimal quantitative cryptanalysis of the permutation-only encryption
schemes with respect to known/chosen plaintext attack.

By combining the Chen system and the Logistic map, a novel permutation-diffusion image encryption algorithm
is proposed in this paper.
To resist the known attacks and achieve better efficiency, two temp-value feedback mechanisms are embedded into
a single permutation-diffusion round.
In the permutation part, we develop different permutation sequences for different plain-images
by means of mapping some information of the plain-image to the generation process of the permutation sequence.
Thus makes the permutation behave in a ``one time pad" manner.
In the diffusion part, another feedback technique is employed to make the
equivalent key generation depend on both the plain-image and the temp-value. By combining the proposed
permutation and diffusion technique, the scheme frustrates the known attacks \cite{Solak:AnalyFridrich:IJBC10,
Chengqing:AnalysisFridrichstructure:IEEECirSys08}.
In addition, we add a reversely-executed diffusion process to make the scheme sensitive to changes of plain-image.


The rest of this paper is organized as follows. Section~\ref{sec:scheme} presents some descriptions
of the prerequisites of the algorithm, such as expanded XOR operation, the temp-value feedback mechanism,
followed by the detailed encryption/decryption procedures.
In Sec.~\ref{sec:securityanalysis}, we evaluate the new scheme via
numerical simulations and comparisons. The last section gives some concluding remarks.

\section{The proposed image encryption algorithm}
\label{sec:scheme}
\subsection{The involved chaotic systems}
Chen system has been widely adopted in many chaotic image encryption algorithms, it can be modeled
by \cite{Chen:3DChaoticCipher:CSF04}
\begin{equation}
\left\{\,
\begin{IEEEeqnarraybox}[][c]{rCl}
\IEEEstrut
$\.{x}$ &=& a(y-x), \\
$\.{y}$ &=& (c-a)x-xz+cy, \\
$\.{z}$ &=& xy-bz,
\IEEEstrut
\end{IEEEeqnarraybox}
\right.
\label{eq:chensystem}
\end{equation}
where $a$, $b$ and $c$ are system parameters. The system is chaotic when $a=35$, $b=3$ and $c \in [20, 28.4]$.
In the proposed encryption algorithm, $c$ is fixed at $28$.

The other chaotic system employed in this encryption algorithm is the Logistic map
\begin{equation*}
y_{n+1} = \mu \cdot y_{n} \cdot (1-y_n),
\end{equation*}
where $y_{n} \in (0, 1)$ and $\mu$ is the control parameter. When $\mu \in (3.5699456, 4)$, the output sequence
is ergodic in the unit interval $(0, 1)$, which makes the Logistic map suitable for pseudorandom
 number generation~\cite{Andrecut:LogisticasRan:IJMPB98}.

\subsection{The expanded $\XOR$ operation}
The expanded XOR ($\eXOR$) operation is introduced to enhance the overall
security level of the scheme. For two inputs $x=\sum_{i=0}^{7}x_{i}\cdot 2^i$ and $r=\sum_{i=0}^{8}r_i\cdot 2^i$,
\begin{equation*}
\eXOR(x,r) = \sum_{i=0}^{7} \NOT ( x_i \oplus r_i\oplus r_{i+1} ) \cdot2^i,
\end{equation*}
where $\NOT(x)$ flips a single bit $x$.
Then one can deduce a property of $\eXOR$ as follows.
\begin{property}
If the equation
\begin{equation*}
\eXOR(x, r)=t
\end{equation*}
holds, then
\begin{equation*}\eXOR(t, r)=x.\end{equation*}
\label{pro:property}
\end{property}
\begin{proof}
This property can be proved by checking every bit of $\eXOR(x, r)$ and $\eXOR(t, r)$, which is given in
table~\ref{tab:eXORresult}.
\begin{table}[!htb]
\centering
\caption{The result of $\NOT ( x_i \oplus r_i\oplus r_{i+1} )$.}
\begin{tabular}{ p{2.4cm}  p{2.4cm}  p{2.4cm}  p{2.4cm}  p{2.4cm}}
\hline\multirow{2}{20pt}{$x_i$}     & \multicolumn{4}{c}{$r_ir_{i+1}$} \\  \cline{2-5}
                                           & $00$    & $01$    &$10$     & $11$ \\
\hline      $0$                             & $1$     & $0$     & $0$    & $1$ \\
\hline      $1$                             & $0$     & $1$     & $1$     &$0$ \\
\hline
\end{tabular}
\label{tab:eXORresult}
\end{table}

\end{proof}

\subsection{The temp-value feedback mechanism}
The core of the proposed encryption algorithm are two feedback techniques called
plaintext feedback in permutation and plaintext/ciphertext feedback in diffusion.
They will be introduced in the following sections.

\subsubsection{Plaintext feedback in permutation}
\label{sec:permutation}
Permutation is a basic component of both traditional and chaos-based encryption algorithms.
In traditional encryption standards, such as DES and AES, fixed permutation tables are employed.
While most chaos-based schemes, such as those in
\cite{Fridrich:ChaoticImageEncryption:IJBC98,Chen:3DChaoticCipher:CSF04,Mao:3Dbakermap:Optic2009,Zhang:ImageCrypt:SCSF07},
use key-dependent permutation. But all of them have one feature in common: the
permutation sequences (tables) are fixed once the secret key is given.
By introducing the plaintext feedback technique, we develop a new approach to design
dynamic permutation order, i.e., different plain-images corresponding to different permutation sequences,
thus makes permutation execute in a ``one time pad'' manner.

Without loss of generality, denote an $8$-bit sequence and its permutated version by $\{a_i\}_{i=1}^{n}$
and $\{a'_i\}_{i=1}^{n}$, respectively. For any $i\in\{1, 2, \cdots, n\}$, permutation only makes change
to the position of $a_i$ while keeps its value unchanged. Thus one has
\begin{align*}
\max(a_i) &= \max(a'_i), \\
\sum_{i=1}^na_i &= \sum_{i=1}^na'_i.
\end{align*}
Set
\begin{equation}
    y_0 =
    \begin{cases}
    0,                                   & \text{if } \max(a_i)=0,\\
    \frac{\sum_{i=1}^n a_i} {n \cdot \max(a_i)} = \frac{\sum_{i=1}^n a'_i} {n \cdot \max(a'_i)}, & \text{otherwise},
    \end{cases}
\label{eq:permutateequation}
\end{equation}
Note that $y_0 =0$ or $y_0 =1$ if and only if $a_i\equiv c$, where $i \in \{1, 2, \cdots, n\}$ and $c$ is a
constant. The permutation can be simply ignored when this situation occurs during encryption.
Then the process to generate a permutation sequence $\{s_i\}_{i=1}^{n}$ from $y_0$ and the Logistic map can be described as follows.
\begin{enumerate}[{step}  1:]
\item Initialize the flag sequence $\{f(k)\}_{k=1}^n$ to $0$ and set $i=1$, $j=1$.
\item{\label{step2}} Calculate $y_j = u \cdot  y_{j-1}  (1- y_{j-1})$ and $y = \lceil y_j \cdot n \rceil$ sequentially.
\item If $f(y)$ is equal to $0$, then set $f(y)=1$, $s_i=y$ and $i=i+1$; otherwise, let $j=j+1$ and go to step~\ref{step2}.
\item If $i<N$, set $j=j+1$ and go to step~\ref{step2}.
\end{enumerate}
Since the process acts like in a Coupon collector \cite[Sec.~3.6]{Motwani:RandomAlgo:Cambridge95} manner, it is concluded that the required complexity to
produce $\{s_i\}_{i=1}^{n}$ is $O(n \log n)$, which outperforms the suggested methods suggestted in \cite{Fouda:fastblock:CNSNS14}, whose complexity
is $O(n+ n \log n)$.


It should be noticed that the inverse permutation shares the same structure of the original one because Eq.~(\ref{eq:permutateequation}) always holds. In the proposed permutation, some information of the
original sequence $\{a_i\}_{i=1}^{n}$ is mapped to the initial value of the nonlinear function.
Then one can generate distinct permutation sequences $\{s_i\}_{i=1}^{n}$ for different plain-images.
Thus makes this kind of permutation tough for the known/chosen plaintext attack proposed in \cite{Chengqing:OptimalofPermutation:SG11, Shujun:quantitativePermutation:SGIC08}.


\subsubsection{Plaintext/ciphertext feedback in diffusion}
\label{sec:diffusion}

The diffusion procedure presented in this scheme is composed of two rounds,
which are denoted by \textit{Diffusion I} and \textit{Diffusion II}, respectively.
In \textit{Diffusion I}, we first calculate the equivalent secret keys according to the gray value of the last
plain-image pixel. Then the current cipher-image pixel can be obtained by combining the current plain-image pixel, the previous cipher-image pixel and the current equivalent keys with the help of the $\eXOR$ operation and modulo $256$ addition.~Namely, each pair of equivalent keys are produced by using the plaintext feedback technique and
every cipher-image pixel is obtained by using the ciphertext feedback technique.~\textit{Diffusion II}
owns the same structure as \textit{Diffusion I} but executes reversely,
thus makes the diffusion process sensitive to changes of plain-image.
Suppose an $8$-bit PRNS $\{x_i\}_{i=0}^{n+3}$ is available, one can compute the cipher-image
pixel sequence $\{c_i\}_{i=1}^n$ from
$\{p_i\}_{i=1}^n$ by:
\begin{itemize}
\item \textit{Diffusion I}:
Set $p_0 = x_n$, $m_0=x_{n+1}$ and $i=1$, obtain $\{m_i\}_{i=1}^n$ by the following steps.
        \begin{itemize}
        \item  Set the initial condition $r_0$ of the Logistic map to
            \begin{equation}
            r_0 =
            \begin{cases}
            (x_{i-1} + 127)/(p_{i-1}+255),     & \text{if } x_{i-1} \leq p_{i-1},\\
            (p_{i-1} + 127)/(x_{i-1}+255), & \text{otherwise}.
            \end{cases}
            \label{eq:generateinitialcondition}
            \end{equation}

        \item   Iterate the Logistic map twice from $r_0$ to obtain $\hat{r}$ and $\hat{r}'$, refresh them        with Eq.~(\ref{eq:refreshr}) and denote the result by $r$ and $r'$, respectively.
            \begin{equation}
            g(x) = \lfloor x \cdot 10^8  \rfloor \bmod 512 ,
            \label{eq:refreshr}
            \end{equation}
            where $\lfloor x  \rfloor$ returns the nearest integer smaller than or equal to $x$.
        \item  Compute $m_i$ by
        \begin{equation}
        m_i = \eXOR(p_i,r) \dotplus \eXOR(m_{i-1},r'),
        \label{eq:computetemp}
        \end{equation}
        where $a \dotplus b = (a+b) \bmod 256$.
        \item Let $i=i+1$, execute the above three steps for $n-1$ times and get $\{m_i\}_{i=1}^n$.
        \end{itemize}

\item \textit{Diffusion II}: Similar to \textit{Diffusion I}, one can obtain $\{c_i\}_{i=1}^n$ by
diffuse $\{m_i\}_{i=1}^{n}$ in reverse order. Set $m_{n+1}= x_{n+3}$, $c_{n+1} = x_{n+2}$, for
$i = n \sim 1$, \textit{Diffusion II} is the same as \textit{Diffusion I} except Eq.~(\ref{eq:generateinitialcondition}) and
Eq.~(\ref{eq:computetemp}) are replaced by
\begin{equation*}
            r_0 =
            \begin{cases}
            (x_{n-i} + 127)/(m_{i+1}+255),     & \text{if } x_{n-i} \leq m_{i+1},\\
            (m_{i+1} + 127)/(x_{n-i}+255), & \text{otherwise},
            \end{cases}
\end{equation*}
and
\begin{equation}
        c_i = \eXOR(m_i,r) \dotplus \eXOR(c_{i+1},r'),
        \label{eq:computecipher}
\end{equation}
respectively.
\end{itemize}

\subsection{Encryption algorithm}
\label{sec:encryption}
Combining the two kinds of temp-value feedback techniques described above, a novel symmetric
gray image encryption scheme is presented in this section. Without loss of generality, one can
scan a plain-image in the raster order and represent it as a one-dimensional $8$-bit sequence
$\{p_i\}_{i=1}^n$. Given the secret key $\bm{K} = (x, y, z, \mu)$, where $(x, y, z)$ is the initial
condition of the Chen system and $\mu \in (3.5699456, 4)$ is the control parameter of the Logistic map, the details of the encryption procedure are described as follows.
\begin{itemize}
\item \textit{Step (1) Initialization}: Implement the Chen system (\ref{eq:chensystem})
by the fourth-order Runge-Kutta method at step size $h=0.001$ iteratively for
$1000+\lceil (n+4)/3 \rceil$ times from $(x, y, z)$. Generate a real number sequence $\{z_i\}_{i=0}^{n+3}$ by
combining the later $\lceil (n+4)/3 \rceil$ approximate states of the Chen system. Then obtain
an $8$-bit PRNS $\{x_i\}_{i=0}^{n+3}$ from $\{z_i\}_{i=0}^{n+3}$ by using
\begin{equation*}
x_i = \lfloor \dec( \left| z_i \right|)\cdot 10^8 \rfloor \bmod 256,
\end{equation*}
where $\left| x \right|$ and  $\dec(x)$ return the absolute value and the fractional part
of $x$, respectively.

\item \textit{Step (2) Permutation}: Permute the plaintext sequence $\{p_i\}_{i=1}^n$ as described in Sec.~\ref{sec:permutation} and denote the sequence after permutation as $\{p'_i\}_{i=1}^n$. The nonlinear
    function adopted here is the Logistic map.

\item \textit{Step (3) Diffusion}: Given PRNS $\{x_i\}_{i=0}^{n+3}$, diffuse $\{p'_i\}_{i=1}^n$
    as described in Sec.~\ref{sec:diffusion} to obtain the ciphertext sequence $\{c_i\}_{i=1}^n$.
\end{itemize}

\subsection{Decryption algorithm}
The decryption procedures are similar to those of encryption except the following modifications:
1)~\textit{Permutation} and \textit{Diffusion} are executed in reverse order,
2)~\textit{Diffusion I} and \textit{Diffusion II} must also be
executed reversely, 3) Referring to property~\ref{pro:property}, Eq.~(\ref{eq:computetemp}) and Eq.~(\ref{eq:computecipher}) should be replaced with
\begin{equation*}
    p_i = \eXOR(m_i \dot{-} \eXOR(m_{i-1},r'), r)
\end{equation*}
and
\begin{equation*}
    m_i = \eXOR(c_i \dot{-} \eXOR(c_{i+1},r'), r),
\end{equation*}
where $a \dot{-} b = (a-b) \bmod 256$.

\section{Security and efficiency analysis}
\label{sec:securityanalysis}
\subsection{Analysis of key space}
It is recommended in \cite{shujun:BasicRrequirement:IJBC06} that the key space of a chaos-based encryption system
should be larger than $2^{100} \approx 10^{30}$ to resist the brute-force attack.
As described in Sec.~\ref{sec:encryption}, the secret key
of the suggested scheme is composed of the double-precision floating-point representation of the initial condition of the Chen
system and the control parameter of the Logistic map, i.e., $\bm{K} = (x, y, z, \mu)$.
The number of significant digits in each parameter is $15$.
Therefore, the key space is $(10^{15})^4 = 10^{60}$, which is much larger
than the recommended size for a secure cipher.

\subsection{Differential analysis}
Differential analysis aims to reveal some information of the (equivalent) secret key of an encryption
algorithm by means of observing how differences in input can affect the resultant output.
To implement the differential
analysis in an image encryption system, the opponent can encrypt two chosen plain-images
with minor modifications, e.g., a slight change in one of the pixels, then compare the encrypted results.
If the minor modification generates significant and unpredictable results in the cipher-image,
this attack will become inefficient.

To give a quantitative description of the one-pixel change on the encrypted results, two common measures
NPCR (number of pixels change rate) and UACI (unified average changing intensity) are used. They are defined
by
\begin{equation*}
\text{NPCR} = \sum_{i,j} \frac{D(i,j)} {H \times W} \times 100\%
\end{equation*}
and
\begin{equation*}
\text{UACI} = \frac{1}{H \times W} \sum_{i,j} \frac{\left| C_1(i,j) - C_2(i,j) \right|}{255}\times 100\%,
\end{equation*}
where $C_1$ and $C_2$ are two $H\times W$ (height$\times$width) cipher-images corresponding to
plain-images different in one single pixel, $C_1(i,j)$/$C_2(i,j)$ is the gray-scale value of $C_1$/$C_2$
at location $(i,j)$ and $D$ is a matrix defined by
\begin{equation*}
            D(i,j) =
            \begin{cases}
            0,     & \text{if } C_1(i,j)=C_2(i,j),\\
            1, & \text{otherwise}.
            \end{cases}
\end{equation*}
It is reported in \cite{Zhu:HyperEnc:OC12} that the expected NPCR and UACI values of a $256$
gray-scale image are $99.6094\%$ and $33.4635\%$, respectively.

The tests of the proposed scheme are carried out as follows. Given a plain-image $P_1$, randomly choose
a location $(i,j)$ and obtain $P_2$ by
\begin{equation*}
            P_2(i,j) =
            \begin{cases}
            P_1(i,j)+1,     & \text{if } P_1(i,j)< 255,\\
            254, & \text{otherwise}.
            \end{cases}
\end{equation*}
Encrypt $P_1$ and $P_2$ and denote the cipher-images by $C_1$ and $C_2$, then one can calculate NPCR and UACI as defined above.
Given the secret key $\bm{K} = (3.0, 4.0, 5.0, 3.999)$, repeat this test $200$ times,
we found that the mean NPCR and UACI values are $99.6041\%$ and $33.4198\%$, respectively,
which are very close to their expectation.

\subsection{Statistical analysis}
\subsubsection{Histograms of encrypted images}
\begin{figure}[!htb]
\centering
\begin{minipage}{\figwidth}
\includegraphics[width=\textwidth]{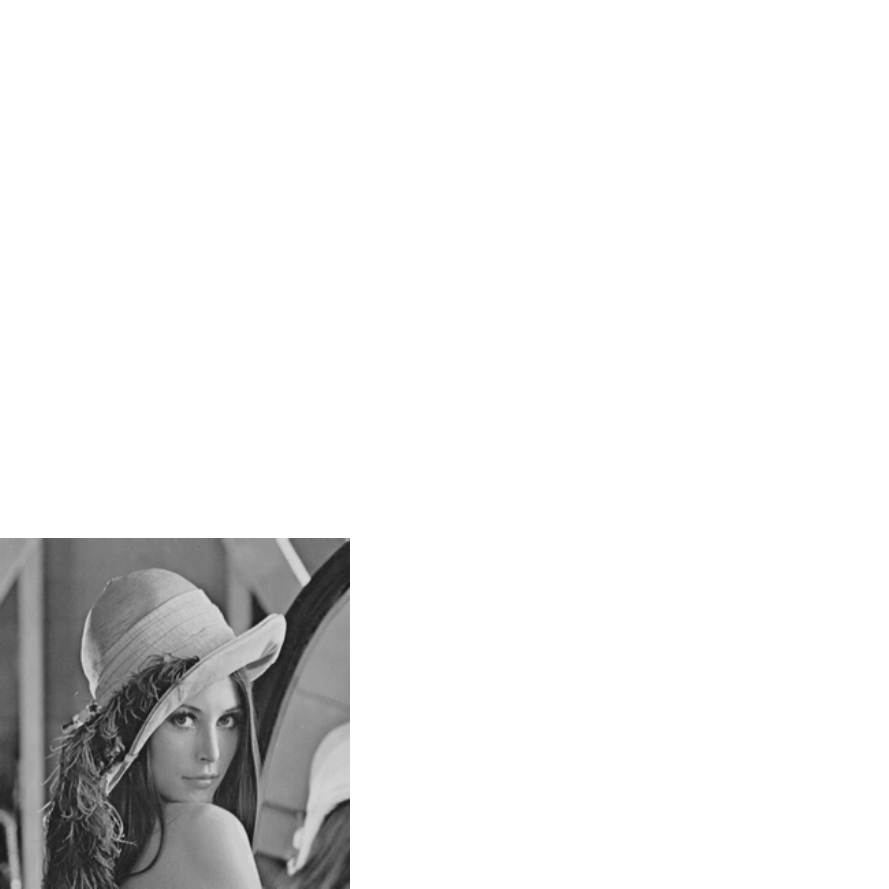}
\center (a)
\end{minipage}
\begin{minipage}{\imagewidth}
\includegraphics[width=\columnwidth]{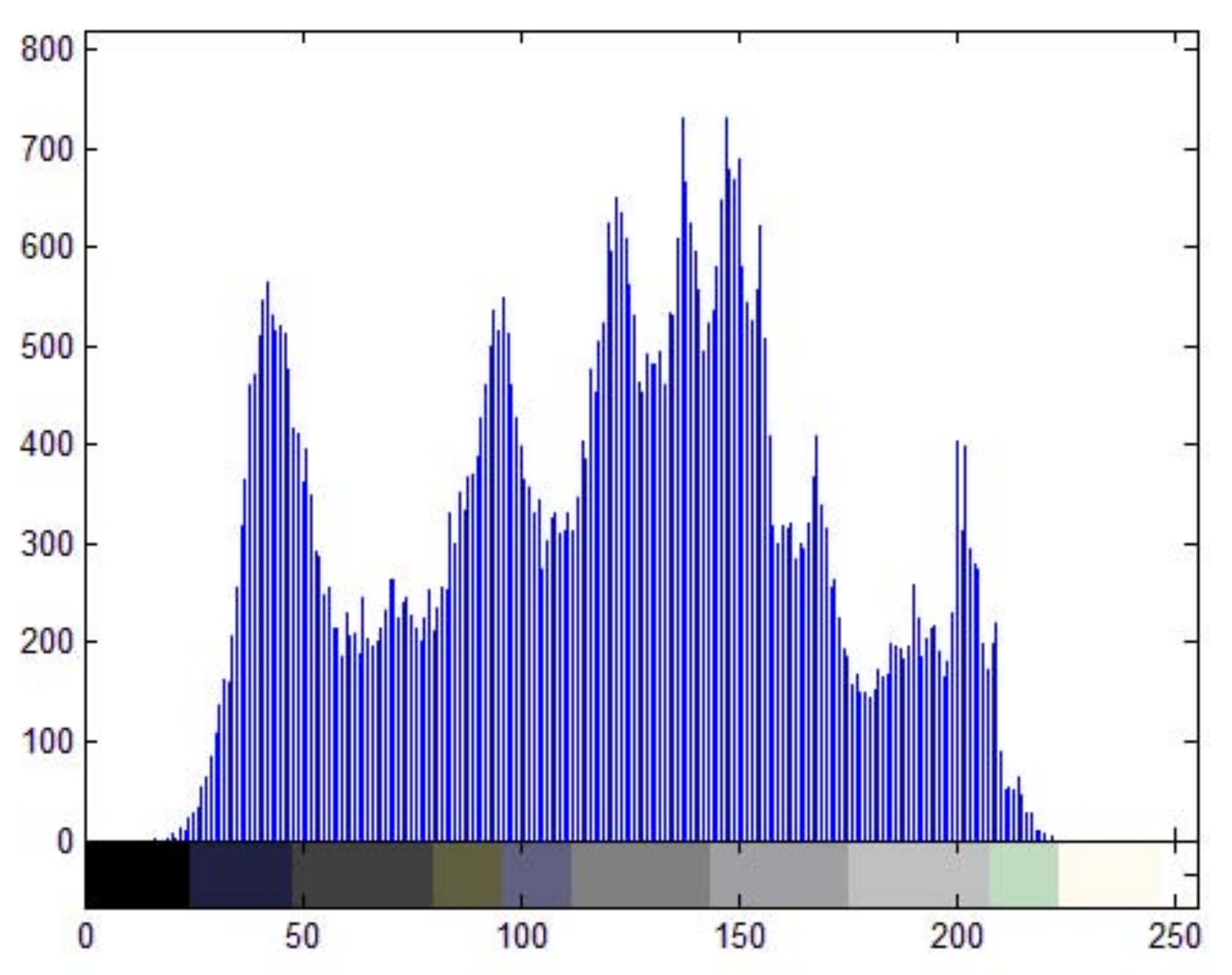}
\center (b)
\end{minipage}\\
\begin{minipage}{\figwidth}
\includegraphics[width=\textwidth]{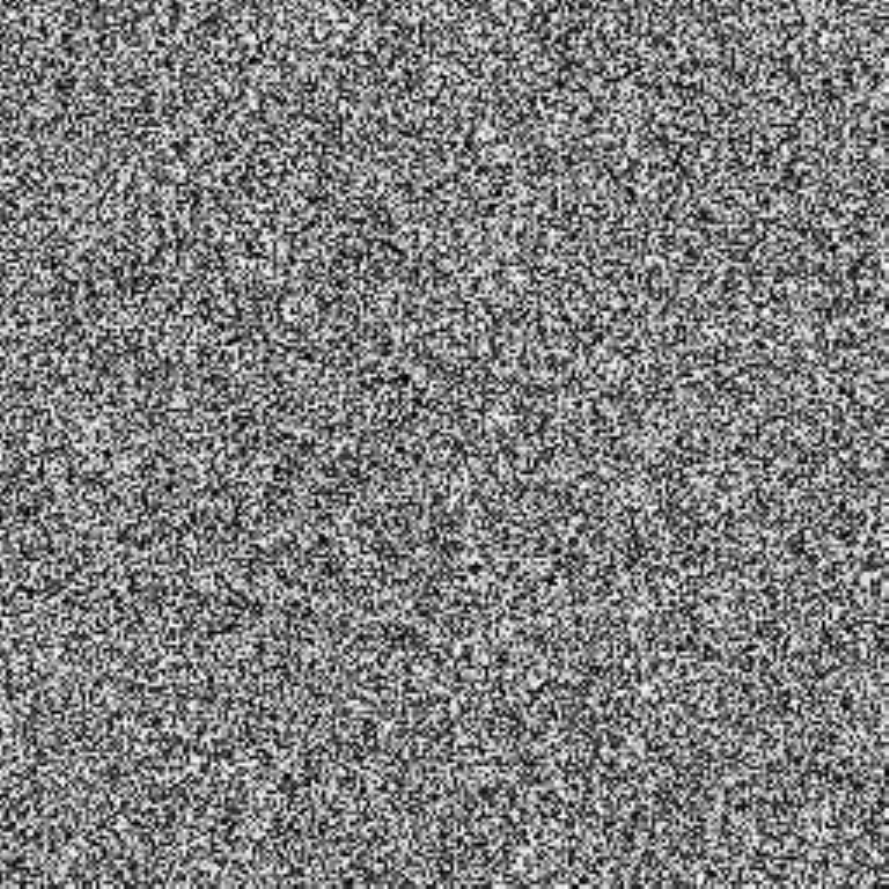}
\center (c)
\end{minipage}
\begin{minipage}{\imagewidth}
\includegraphics[width=\columnwidth]{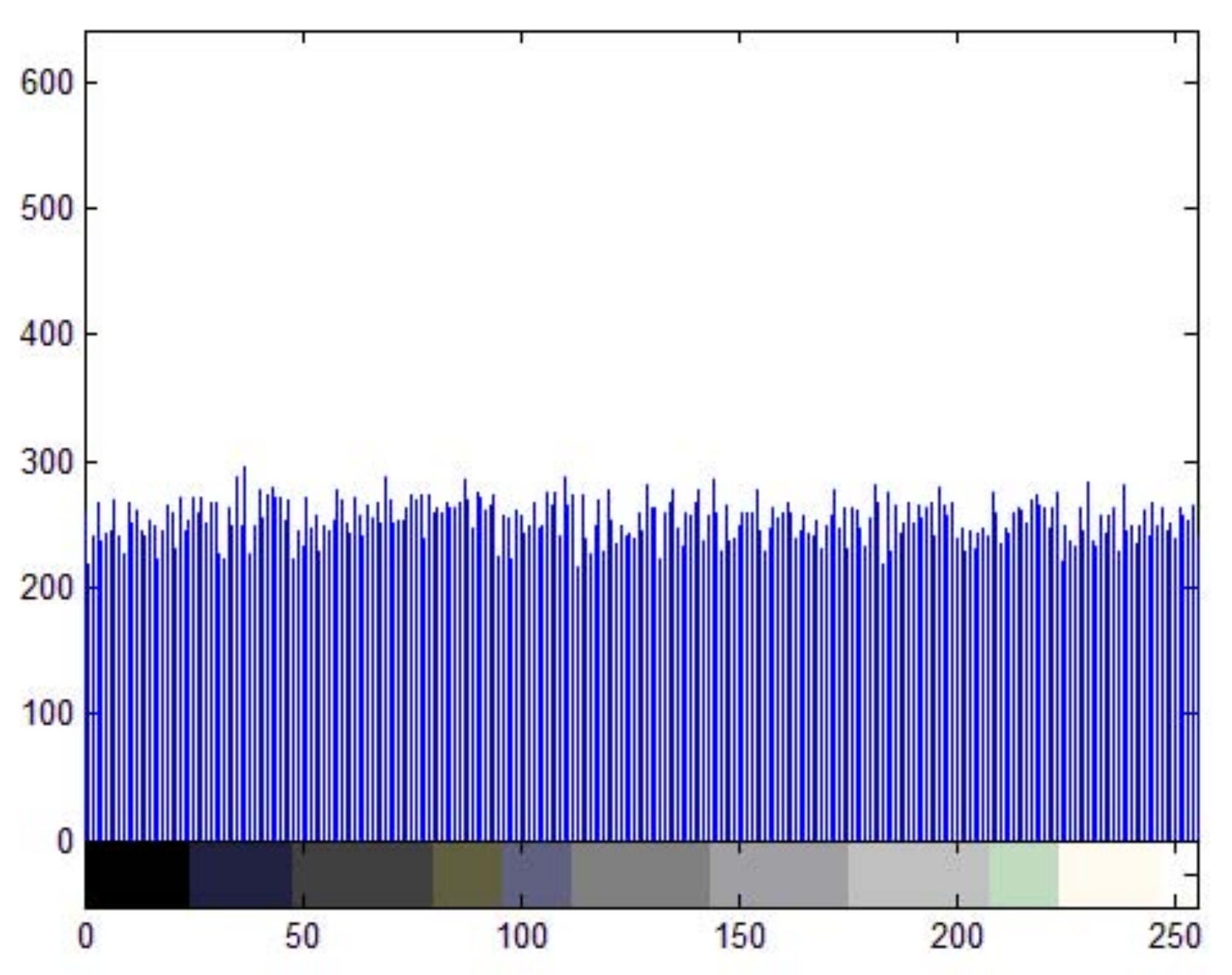}
\center (d)
\end{minipage}
\caption{Histograms of plain-image ``Lenna" and its corresponding cipher-image:
(a) plain-image ``Lenna";
(b) Histogram of ``Lenna";
(c) the cipher-image of ``Lenna";
(d) Histogram of the cipher-image.}
\label{figure:histograms}
\end{figure}

In Figure~\ref{figure:histograms}, we give a typical example showing histograms of the plain-image and
the corresponding cipher-image. As shown in Fig.~\ref{figure:histograms}(d), all the gray-scale values
of the cipher-image of ``Lenna" are distributed uniformly over the interval $[0,255]$, which is significantly
different from the original distribution shown in Fig.~\ref{figure:histograms}(b).

\subsubsection{Information entropy}
The histogram gives a visual sense of the pixel distribution in the cipher-image, while information entropy
can offer a measure to quantify the random-looking distribution. The most commonly used information entropy is
the Shannon entropy~\cite{Shannon:Entropy:BSTJ49}, which is regarded as the vital feature of randomness.
For a message source with $2^N$ symbols $m_i$, its Shannon entropy, $H(m)$, is defined as follows:
\begin{equation*}
H(m) = - \sum_{i=0}^{2^N -1} Pr(m_i) \cdot \log_2[Pr(m_i)],
\end{equation*}
where $Pr(m_i)$ is the probability of occurence of the symbol $m_i$ which is determined by the source. It is easy to prove
that a random gray-scale image with uniformly distributed pixels over the
interval $[0, 255]$ can achieve the ideal Shannon entropy $8$. Set the secret key $\bm{K} = (2.0, 3.0, 4.0, 3.9876)$,
and get the cipher-image of ``Lenna", ``Baboon" and ``Pepper" of the same size $256\times 256$.
The Shannon entropy of the three cipher-images are $H_{\text{Lenna}}=7.9973$, $H_{\text{Baboon}}=7.9971$
and $H_{\text{Pepper}}=7.9969$, which are very close to the ideal value $8$.

\subsubsection{Correlation of adjacent pixels}
Adjacent pixels having high correlation
is an intrinsic characteristic of digital images without compression.
An effective image encryption algorithm should be able to remove this kind of relationship. To test the correlation
between horizontally, vertically and diagonally adjacent pixels,
we calculate the correlation coefficient in each direction by
\begin{equation*}
cov(x,y)=\frac{ \frac{1}{N}\cdot \sum_{i=1}^N (x_i- \bar{x}) \cdot (y_i- \bar{y} ) }
{\sqrt{(\frac {1}{N} \sum_{i=1}^N(x_i- \bar{x})^2)\cdot (\frac {1}{N} \sum_{i=1}^N( y_i- \bar{y} )^2)}},
\end{equation*}
where $\bar{x}=\frac{1}{N} \cdot \sum_{i=1}^N x_i$, $\bar{y}= \frac{1}{N} \cdot \sum_{i=1}^N y_i$, $(x_i, y_i)$ is
the $i$-th pair of adjacent pixels in the same direction and $N$ is the total number of pixel pairs.
The result of the correlation coefficients along the three directions of the plain-image ``Lenna" and its corresponding
cipher-image under secret key $\bm{K} = (5.0, 3.0, 4.0, 3.999)$ are listed in Table~\ref{tab:correaltionship}.
It is clear that the correlation is almost reduced to $0$ after encryption.
\begin{table}[!htb]
\centering \caption{Correlation coefficients between the plain-image ``Lenna" and the corresponding cipher-image.}
 \begin{tabular}{ p{4.2cm}  p{4.2cm}  p{4.2cm}}
\hline   Direction    & Plain-image ``Lenna"    & After encryption     \\
\hline      horizontal  & $0.93903$     & $0.00350$     \\
\hline      vertical    & $0.96812$     & $0.00247$    \\
\hline      diagonal    & $0.91352$     & $0.00107$     \\
\hline
\end{tabular}
\label{tab:correaltionship}
\end{table}

\subsection{Speed performance}
We also test the average encryption (decryption) speed of the proposed scheme on a $3.2$GHz Intel
Pentium Dual Core CPU with $2$GB RAM using VC compiler.
Table~\ref{tab:runspeed} gives the average speed of the
proposed scheme with respect to images of various size.
For comparison, the running speed of the algorithm suggested in \cite{Chen:3DChaoticCipher:CSF04} and DES
are also listed. The results indicate that our scheme operates in a fast manner.

\begin{table}[!htb]
\centering
\caption{Speed performance of the the proposed scheme, the scheme in \cite{Chen:3DChaoticCipher:CSF04} and
DES algorithm.}
\begin{tabular}{ p{2cm}  p{3.6cm}  p{3cm}  p{3.4cm} }
\hline  {Image size (pixels)} &{The proposed scheme (ms)}    &{The  algorithm in \cite{Chen:3DChaoticCipher:CSF04} (ms)}
                                                            &{DES algorithm (ms)}\\
\hline      $256\times 256$     & $22$     & $60$           & $28$                \\
\hline      $512\times 512$     & $98$     & $273$          & $110$                \\
\hline      $1024\times 1024$   & $415$    & $1180$         & $445$               \\
\hline
\end{tabular}
\label{tab:runspeed}
\end{table}

\section{Conclusion}
This paper presents a simple but secure chaotic cipher for gray images by
improving the familiar permutation-diffusion structure.
As the plaintext feedback technique is used during permutation, one can develop different permutation
sequences for different plain-images, which makes the scheme immune to known/chosen plaintext attack.
The diffusion procedure is better than the traditional one by employing a nonlinear
operator $\eXOR$ and generating equivalent key stream dynamically.
Experimental tests demonstrate that the scheme possesses large key space, high security
and good encryption (decryption) speed. Thus, it may serve as a candidate for real-time image encryption application.

\section*{Acknowledgement}
This research was supported by the technology project of State Grid of China (No.~XX17201200048).

\bibliographystyle{elsarticle-num}
\bibliography{CNSNS}
\end{document}